# Nonequilibrium and nonlinear kinetics as key determinants for bistability in fission yeast G2-M transition

De Zhao[1,2*], Teng Wang[1*], Jian Zhao[2,3*], Dianjie Li[1*], Zhili Lin[1], Zeyan Chen[1], Qi Ouyang[1], Hong Qian[4†], Yu V. Fu[2,3†], Fangting Li[1†]

[1]School of Physics, Center for Quantitative Biology, Peking University, Beijing 100871, China.

[2]State Key Laboratory of Microbial Resources, Institute of Microbiology, Chinese Academy of Sciences, Beijing 100101, China.

[3] Savaid Medical School, University of Chinese Academy of Sciences, Beijing 101408, China.

[4] Department of Applied Mathematics, University of Washington, Seattle, Washington 98195, U.S.A.

*These authors contributed equally to this work.

† Corresponding authors. E-mail: hqian@u.washington.edu; fuyu@im.ac.cn; lft@pku.edu.cn

## Abstract

A living cell is an open, nonequilibrium biochemical system where ATP hydrolysis serves as the energy source for a wide range of intracellular processes, possibly including the assurance for decision-making. In the fission yeast cell cycle, the transition from G2 to M phase is driven by the activation of Cdc13/Cdc2 and Cdc25 and the deactivation of Wee1 through phosphorylation-dephosphorylation cycles with feedback loops. Here, we present a kinetic description of the G2-M circuit which reveals that both cellular ATP level and ATP hydrolysis free energy critically control Cdc2 activation. Using fission yeast nucleoplasmic extract (YNPE), we experimentally verify that increased ATP level drives the activation of Cdc2 which exhibits bistability and hysteresis in response to changes in cellular ATP level and ATP hydrolysis energy. These findings suggest that cellular ATP level and ATP hydrolysis energy are determinants of the bistability and robustness of Cdc2 activation during G2-M transition.

**Significance**

Living cells operate far from thermodynamic equilibrium and utilize ATP hydrolysis to power various cellular processes. Here, we address how cellular ATP level and ATP hydrolysis free energy influence biochemical regulation in fission yeast cell cycle to determine the fundamental connection between cellular fate decision and nonequilibrium energetics. Our results showed that the CDK activation circuit of G2/M transition in fission yeast cell cycle exhibits bistability and hysteresis by a mechanism known as saddle-node bifurcation, but this only occurs when the yeast cell is far from equilibrium (large free energy). This switching mechanism is globally robust against cellular fluctuations in both phosphorylation energy level and protein concentration. So, the free energy in living systems definitely affects decision-making and information processes.

## Introduction

Biochemical processes in the living cell operate far from equilibrium with biochemical circuits and networks controlling various cellular functions. E. Schrödinger first suggested that living organisms require negative entropy flux to create and maintain order[1]. I. Prigogine and colleagues proposed that self-organization in nonequilibrium systems produces orders called dissipative structures[2]. From a thermodynamic perspective, nonequilibrium nature or free energy input is indispensable for living organisms[3, 4]. Beyond the traditional view that free energy is consumed to carry out processes like biosynthesis, ionic pumping, or mechanical movements[3], the recent phosphorylation energy hypothesis proposed that free energy is necessary for information processing and decision-making in living cells[4]. Without adequate free energy input, cells would not be biologically functional, no matter how well-designed are the biochemical networks. Theoretical studies have shown that free energy is a decisive factor, endowing biochemical reactions with the characteristics of ultrasensitivity, bistability or oscillation. As a consequence, biochemical molecules can execute information processing, state transition and decision-making in living cells[5, 6, 7, 8, 9, 10, 11, 12, 13, 14, 15]. So far, however, no direct experimental evidence has shown that free energy actually plays a role in the cellular state transition of eukaryotes.

In order to correctly produce two daughter cells from a single mother cell, the eukaryotic cell cycle is divided into four phases. DNA is replicated in the S phase, and the cell undergoes mitosis in the M phase. The S and M phases are separated by two gaps called G1 phase and G2 phase, respectively. Phase transitions in the cell cycle are well-characterized biological decision-making processes, and they are controlled by biochemical reaction networks with complex feedback loops[16, 17, 18]. In cell cycle progression, cyclins, i.e., proteins whose concentration periodically rises and falls to regulate cell cycle processes, activate cyclin-dependent kinase (CDK) and thereby drive the transition from one phase to the next[19, 20].

In the fission yeast Schizosaccharomyces pombe, transition from G2 to M phase is promoted by the activation of CDK, a mechanism that is highly conserved from yeast to humans[16, 19, 20, 21, 22]. Fission yeast CDK, Cdc2, forms a kinase complex with Cdc13, a B-type cyclin. The Cdc13/Cdc2 complex is activated by Cdc25 and inhibited by Wee1[22, 23, 24] (Fig. 1a). Among these three components, both positive and double-negative feedbacks exist whereby active Cdc13/Cdc2 complex activates the phosphatase activity of Cdc25 and inactivates the kinase activity of Wee1, both through phosphorylation on multiple sites[25, 26, 27, 28]. Phosphorylation-dephosphorylation (PdP) cycles of Cdc13/Cdc2 complex, Cdc25 and Wee1 are actively driven by the free energy generated from ATP hydrolysis[25, 26]. Inside a living cell, free energy generated from the hydrolysis of one mole of ATP, also called phosphorylation energy, is defined as $\Delta G = RT \ln \gamma$, where $\gamma = [ATP]/(K_{eq}[ADP][Pi])$ and $K_{eq}$ is the equilibrium constant for ATP synthesis[3,4]. Under physiological conditions in living cells, γ and $\Delta G$ are approximately $10^{10}$ and 57 kJ/mol, respectively[3, 4]. $\Delta G$ , which is the chemical driving force for PdP cycles, connects with chemical reaction fluxes and dissipation of the living cell. Given that free energy $\Delta G$ quantifies how far away a living cell is from a thermochemical equilibrium state, a positive $\Delta G$ distinguishes living state from dead[3, 4].

It has been widely accepted that strong nonlinearity generated by feedbacks results in the bistability of a system[16, 29, 30, 31]. The biochemical switch with bistability and hysteresis in CDK activation, which is mainly caused by saddle-node bifurcation, provides a stable and robust activated CDK state against the fluctuation of cyclin level[32], thus benefiting irreversible transitions in the cell cycle process[16]. Cdc2 activation driven by cyclin B accumulation has been experimentally demonstrated to be bistable in oocyte extracts[29, 30]. It is important to further quantify how intracellular nonequilibrium environment and free energy jointly influence cellular decision-making processes and phase transitions, such as the activation of CDK in fission yeast G2-M transition.

In this work, we want to understand how free energy drives G2-M transition through activating the mitotic regulator CDK in fission yeast. To do this, we first constructed a thermodynamically valid kinetic model of the G2-M circuit. Theoretical analysis clearly shows that both cellular ATP level and phosphorylation energy are critical in controlling the dynamics and bistable nature of Cdc2 activation. Furthermore, to verify our theoretical predictions, we employed fission yeast nucleoplasmic extracts (YNPE) and experimentally monitored Cdc2 activation with different ATP and ADP inputs, together with the measurement of ATP, ADP and phosphorylation energy levels in this system. Phosphorylation energy is experimentally determined by the cellular ATP/ADP ratio. We found that increasing Cdc13 or ATP drives the activation of Cdc2 in YNPE. Cdc2 activity exhibits bistability and hysteresis in response to changes of cellular ATP and phosphorylation energy. On the other hand, we found that increasing Cdc13 reduces the ATP threshold required for Cdc2 activation. Our results show that the CDK activation circuit in G2-M transition exhibits bistability and hysteresis by a mechanism known as saddle-node bifurcation, but this only occurs when the yeast cell is far from equilibrium with a large positive $\Delta G$. This switching mechanism is globally stable and robust against cellular fluctuations in ATP concentration, phosphorylation energy level and cyclin protein concentration where phosphorylation energy level can be influenced by cellular metabolism.

## Model and results

### Mathematical model analysis predicts that Cdc13, cellular ATP level and phosphorylation energy drive Cdc2 activation

To reveal the roles of phosphorylation energy $\Delta G$ and cellular ATP level in G2-M transition in the fission yeast cell cycle, we developed a theoretical analysis of the activation of CDK in fission yeast G2-M transition based on the Cdc13/Cdc2-Wee1-Cdc25 circuit, as shown in Fig. 1a. The activation-inactivation of Cdc13/Cdc2 is represented by chemical reaction functions as follows:

$$Cdc13/Cdc2^{A}+Wee1^{A}+ATP \underset{k_{-1}}{\overset{k_{+1}}{\rightleftarrows}} Cdc13/Cdc2^{A}\cdot Wee1^{A}\cdot ATP \underset{k_{-2}}{\overset{k_{+2}}{\rightleftarrows}} Cdc13/Cdc2^{In}+Wee1^{A}+ADP,$$

$$Cdc13/Cdc2^{In}+Cdc25^{A} \underset{k_{-3}}{\overset{k_{+3}}{\rightleftarrows}} Cdc13/Cdc2^{In}\cdot Cdc25^{A} \underset{k_{-4}}{\overset{k_{+4}}{\rightleftarrows}} Cdc13/Cdc2^{A}+Cdc25^{A}+Pi,$$

where $k_{\pm 2}, k_{\pm 1}, k_{\pm 3}$ and $k_{\pm 4}$ are the kinetic rate constants, and the superscript 'A' represents 'Active' and 'In' represents 'Inactive'. $Cdc13/Cdc2^{In}$ is the inactive state and the phosphorylated form of the Cdc13/Cdc2 complex, while $Cdc13/Cdc2^{A}$ is the active state of the Cdc13/Cdc2 complex. We assumed that Tyr15 was the only site involved in activating the Cdc13/Cdc2 complex[21, 22]. The regulation of Wee1 and Cdc25 by Cdc13/Cdc2 is respectively illustrated in Fig. S1 in Supplementary Information (SI), and the phosphorylation process of Wee1 and Cdc25 by Cdc13/Cdc2 kinase involves multiple sites[27, 28, 33, 34]; $[Wee1^{A}]$ and $[Cdc25^{A}]$ represent the concentrations of active Wee1 and Cdc25, respectively.

Since Cdc2 concentration exceeds that of Cdc13 in yeast cells[35, 36] and since Cdc13 combines with Cdc2 with high affinity, the total concentration of Cdc13 can be represented by $[Cdc13/Cdc2^{T}]$. $[Cdc13/Cdc2^{In}]$ and $[Cdc13/Cdc2^{A}]$ respectively represent the concentration of inactive and active Cdc13/Cdc2.

Thus, the dynamic equation of $[Cdc13/Cdc2^{A}]$ in the fission yeast G2-M system can be described by

$$\frac{d[Cdc13/Cdc2^{\mathrm{A}}]}{dt} = J_p - J_w,$$

where $J_p$ and $J_w$ are the chemical fluxes generated by Cdc25 and Wee1, respectively, and $[ATP]$, $[ADP]$ and $[Pi]$ are the cellular concentrations of ATP, ADP and Pi, respectively, in fission yeast cells.

We set $a_+ = \frac{k_{+1}k_{+2}}{k_{-1}+k_{+2}}$, $a_- = \frac{k_{-1}k_{-2}}{k_{-1}+k_{+2}}$, $b_+ = \frac{k_{+3}k_{+4}}{k_{-3}+k_{+4}}$, $b_- = \frac{k_{-3}k_{-4}}{k_{-3}+k_{+4}}$, then have

$$J_w = \frac{\{a_+[ATP][Cdc13/Cdc2^{A}] - a_-[ADP]([Cdc13/Cdc2^{T}] - [Cdc13/Cdc2^{A}])\}[Wee1^{A}]}{1 + \frac{a_+[ATP][Cdc13/Cdc2^{A}]}{k_{+2}} + \frac{a_-[ADP]([Cdc13/Cdc2^{T}] - [Cdc13/Cdc2^{A}])}{k_{-1}}},$$

$$J_p = \frac{\{b_+([Cdc13/Cdc2^{T}] - [Cdc13/Cdc2^{A}]) - b_-[Pi][Cdc13/Cdc2^{A}]\}[Cdc25^{A}]}{1 + \frac{b_+([Cdc13/Cdc2^{T}] - [Cdc13/Cdc2^{A}])}{k_{+4}} + \frac{\mathrm{b}_-[Pi][Cdc13/Cdc2^{A}]}{k_{-3}}}.$$

So far, the previously existing mathematical descriptions for these biochemical networks[18, 33], which are based on the assumption of irreversible PdP cycles, have been thermodynamically incomplete and have ignored the role of $\Delta G$ and cellular ATP level. In our thermodynamically valid kinetic model, both phosphorylation by kinase and dephosphorylation by phosphatase are described as reversible with a sufficiently large ratio of forward and reverse reaction rates.

Then our model gives thermodynamic predictions on the activation of Cdc13/Cdc2 in G2-M transition; in particular, we obtain the steady-state concentration of active Cdc13/Cdc2 as a function of $[Cdc13/Cdc2^T]$, phosphorylation energy $\Delta G$ and $[ATP]$:

$$[Cdc13/Cdc2^A]=[Cdc13/Cdc2^T]\frac{\theta+\mu}{\theta+\mu+\theta/(\mu\gamma)+1},$$

where $\theta=\frac{a_+[ATP][Wee1^A]}{b_+[Cdc25^A]}$, $\mu=\frac{b_-[Pi]}{b_+}$, and $\gamma=\frac{a_+b_+[ATP]}{a_-b_-[ADP][Pi]}$.

$[Wee1^A]$ and $[Cdc25^A]$ are dependent on $[Cdc13/Cdc2^A]$ and $[ATP]$, as well as free energy level (see SI, section 1.2.2 for details).

Phosphorylation energy level ΔG can be expressed as $\Delta G{=}RT\ln\gamma$. Under physiological conditions of living cells, $\Delta G$ and γ are approximately 57 kJ/mol and $10^{3,4}$, respectively. Especially, when the G2-M system reaches equilibrium state, namely $\Delta G=0$ and $\gamma=1$, the concentration of active Cdc13/Cdc2 becomes $\frac{\mu}{1+\mu}[Cdc13/Cdc2^T]$, which is independent of both $[Wee1^A]$ and $[Cdc25^A]$.

G2-M transition contains multiple PdP cycles, including Cdc13/Cdc2, Wee1 and Cdc25, and then we obtain the following results. Phosphorylation energy $\Delta G$, together with cellular ATP and cyclin Cdc13 levels, governs the activities of CDK kinase Cdc13/Cdc2, Wee1 and Cdc25. Therefore, only when the system is at nonequilibrium state with $\Delta G > 0$ is the nonlinear activation of Cdc13/Cdc2 through Wee1 and Cdc25 able to function. The key factors that govern the activation of Cdc13/Cdc2 are cyclin Cdc13, free energy $\Delta G$ and cellular ATP level.

**Cdc2 activation is theoretically predicted to exhibit bistability and hysteresis controlled by cyclin, ATP and phosphorylation energy $\Delta G$.**

In this section, we first discussed the quantitative behavior of G2-M transition in fission yeast under physiological conditions. Next, we considered the living cells as ideal biochemical reaction systems to investigate G2-M transition under different cellular ATP concentrations, Cdc13 concentrations and $\Delta G$ levels, in particular much lower $\Delta G$ levels than physiological values.

In living fission yeast cells, a high concentration of ATP (3 ~ 4 mM)[37] and low concentration of ADP (0.1 ~ 1.4 mM)[4, 38] are present. The concentration of cellular Pi is estimated between 1 and 10 mM without erratic fluctuation *in vivo*[4]. For the sake of simplification, we assume that cellular Pi concentration $[Pi]$ does not change during G2-M transition and set $[Pi]=1\ \mathrm{mM}$ as a fixed value in our G2-M transition model[4]. Based on these facts, we regard $[ATP]$ and free energy $\Delta G$ as system parameters, and $[ADP]$ can be determined by $\Delta G$ and $[ATP]$. Meanwhile, we also assumed that the phosphorylation processes of Wee1 and Cdc25 by Cdc13/Cdc2 kinase are five-site phosphorylation processes based on biological studies[17, 22, 23, 24, 25, 27, 28, 33, 34]. In our simulation, we set $a_+ = 2.3\times10^{-7} nM^{-2}s^{-1}$ , $a_- = 10^{-10} nM^{-2}s^{-1}$ , $b_+ = 1\ nM^{-1}s^{-1}$ , and $b_- = 10^{-11} nM^{-2}s^{-1}$ to ensure that the bistability of Cdc2 exists under physiological conditions. Parameter sensitivity analysis is performed and discussed in section 1.3 of SI, and parameter values are listed in Table S1.

As shown in Fig. 1b and Fig. 1c, we simulate the activation of Cdc2 in fission yeast G2-M transition by increasing the total concentration of cyclin B, $[Cdc13/Cdc2^T]$, under physiological condition, as well as various $\Delta G$ levels. In Fig. 1b, the lower branch, with low $[Cdc13/Cdc2^T]$ and low Cdc13/Cdc2 activity, represents the state of G2 phase (G2 state), while the upper branch, with both high $[Cdc13/Cdc2^T]$ and high Cdc13/Cdc2 activity, represents the state of M phase (M state). To simulate G2-M transition under physiological condition, we set $\Delta G$=56.8 kJ/mol , $[ATP]$=4.3 mM and $[Pi]$=1 mM, then $[ADP]$=0.1 mM; we assume that $[Cdc13/Cdc2^T]$=60 nM in the G2 phase state[39]. The red line in Fig. 1b represents the widely accepted quantitative behavior of G2-M transition in physiological condition, which is a toggle switch between G2 state and M state driven by Cdc13 accumulation[29]. With the growth of yeast cell, the $[Cdc13/Cdc2^T]$ is accumulated; once the $[Cdc13/Cdc2^T]$ level is increased over the threshold (about 79.4 nM in our model), the activity of Cdc13/Cdc2 increases sharply to a level corresponding to M state. Suppose that the initial state of yeast cell is the M state (high Cdc13/Cdc2 activity in the upper branch of Fig. 1b), then according to our predicted results, the activity of Cdc13/Cdc2 will drop sharply back to the G2 state, but only when $[Cdc13/Cdc2^T]$ is lower than another threshold (about 48.2 nM in our model). We call this hypothetical M-G2 transition process as the 'coming-down' process, whereas the G2-M transition is noted as the 'going-up' process. Our results suggest that bistability and hysteresis exist in cyclin B-driving Cdc2 activation.

When $\Delta G > 34.6\,\text{kJ/mol}$, with $[ATP]{=}4.3\text{ mM}$ and $[Pi]{=}1\text{ mM}$, the G2-M transition behaves with similar bistability for cyclin B-driving Cdc2 activation. However, when $\Delta G \leq 34.6\,\text{kJ/mol}$, G2-M transition behaves with monostability to the change of cyclin B such that the activation of Cdc2 is sensitive to the change of cyclin B during G2-M transition (Fig. 1b).

In Fig. 1c, the two-dimensional parameter space (phase diagram) illustrates how the two driving factors, $[Cdc13/Cdc2^T]$ and $\Delta G$, synergistically determine the activation of Cdc2 under $[ATP]{=}4.3\text{ mM}$. The U-shaped curve area represents the bistable area. From the low Cdc2 activity, trajectories passing through the upper boundary that transverse the bistable region represent the G2-M transition and are accompanied by sharply increasing Cdc2 activity. When $34.6\,\text{kJ/mol} < \Delta G < 52.2\,\text{kJ/mol}$, the higher $\Delta G$ needs a larger $[Cdc13/Cdc2^T]$ to pass the G2-M transition. When $\Delta G > 52.2\,\text{kJ/mol}$, the threshold of $[Cdc13/Cdc2^T]$ for G2-M transition does not change obviously with $\Delta G$ level.

We simulate ATP-driving Cdc2 activation in fission yeast G2-M transition, as illustrated in Fig.1d, Fig.1e and Fig.1f. The red curve in Fig. 1d shows the physiological condition with $\Delta G = 56.8\ \text{kJ/mol}$ and fixed $[Pi] = 1\ \text{mM}$. Suppose yeast cells start from G2 state with $[Cdc13/Cdc2^T] = 60\ \text{nM}$, if we increase $[ATP]$ level, then the system will move to M state once $[ATP]$ is larger than 4.67 mM. On the other hand, the threshold of $[ATP]$ for M-G2 transition is 3.33 mM. Thus, similar to cyclin B-driving Cdc2 activation, ATP-driving activation also exhibits bistability and hysteresis.

Taking the above results together, the bistability and hysteresis of Cdc2 activation caused by saddle-node bifurcation under high $\Delta G$ level ensure the stability of high-level Cdc2 in M phase against the changes of both Cdc13 and ATP from environmental and cellular fluctuations. However, in the monostable range under the lower $\Delta G$ level, the environmental or cellular changes or fluctuations of Cdc13 and ATP may cause a large change of Cdc2 activities in G2-M transition, especially in the graded or sharp Cdc2 change range, which could result in multiple entries of M phase and the instability of genetic hereditary information. In the case of transcritical bifurcation in Cdc2 activation, similar phenomena would happen near the bifurcation point.

We simulated ATP-driving Cdc2 activation under different levels of both $\Delta G$ and $[Cdc13/Cdc2^T]$. The phase diagrams in Fig. 1f illustrate how the three driving factors, $[ATP]$, $\Delta G$ and $[Cdc13/Cdc2^T]$, synergistically determine the activation of Cdc2. The U-shaped curve area represents the bistable area. For example, set $[Cdc13/Cdc2^T]$=60 nM; only when $\Delta G > 33.7$ kJ/mol will $[ATP]$-driving Cdc2 activation exhibit bistability and hysteresis. In the high $\Delta G$ range ($\Delta G > 45.7$ kJ/mol), the ATP thresholds for Cdc2 activation basically remain constant; we denote this as the ATP-constant U-shaped curve range. Furthermore, a critical value of $\Delta G$ (33.7 kJ/mol in our model) is set for the transition between monostability and bistability. In subsection 1.2.3 of SI, we investigate how to increase the critical value of $\Delta G$ to the physiological value (~57 kJ/mol) by changing the kinetic reaction rates in our model. We find that a possible method is to decrease the rate constant of Cdc25-regulated Cdc13/Cdc2 inactivation ($k_{-4}$) by nearly 10,000 times. From a biochemical point of view, this is difficult to achieve. In addition, this may suggest that the G2-M circuit is working in the ATP-constant U-shaped curve range to avoid the fluctuation of cellular $\Delta G$ level.

Under different $[Cdc13/Cdc2^T]$ levels, we observe different U-shaped curves, and our results show that only when $[Cdc13/Cdc2^T] > 35$ nM does the $[ATP]$-driving Cdc2 activation exhibit bistability and hysteresis.

We illustrate under physiological level ($\Delta G$=56.8 kJ/mol) in Fig.1g, in addition to other $\Delta G$ levels (Fig. S4a and Fig. S12b in SI), how $[Cdc13/Cdc2^T]$ levels influence the thresholds of $[ATP]$ for G2-M transition (right branch) and M-G2 transition (left branch), respectively. For G2-M and M-G2 transition in the ATP-constant U curve $\Delta G$ range, higher $[Cdc13/Cdc2^T]$ levels require lower $[ATP]$ levels. When $\Delta G < 45.7\ \mathrm{kJ/mol}$, which is lower than the ATP-constant U curve $\Delta G$ range, such as $\Delta G = 33 \text{ or } 36\ \mathrm{kJ/mol}$, lower $[ATP]$ levels are similarly required for higher $[Cdc13/Cdc2^T]$ levels (Fig. S4a and Fig. S12b in SI); however, under a certain $[Cdc13/Cdc2^T]$ level, lower $[ATP]$ threshold levels are required to decrease $\Delta G$ levels with corresponding bistability range (Fig. 1f).

Taken together, we conclude that the cellular decision-making processes in fission yeast G2-M transition are regulated by the free energy of ATP hydrolysis and cellular ATP level and that bistability and hysteresis exist in Cdc2 activation. In a noisy cell, bistability can prevent the high Cdc2 activity of M state from falling to the low Cdc2 activity of G2 state owing to fluctuations in cyclin B and cellular ATP or ADP levels, resulting in the stabilization of Cdc2 activation during G2-M transition.

**Cdc13 promotes the activation of Cdc2 in the YNPE system.**

In order to verify the above theoretical predictions experimentally, especially the roles of ATP and $\Delta G$ in Cdc2 activation, we sought to establish a reliable and operable experimental system. In living yeast cells, it is hard to quantify either $\Delta G$ level or Cdc2 activity *in vivo*. Moreover, owing to the existence of mitochondria and glycolysis, it is unlikely to precisely and independently manipulate ATP and $\Delta G$ levels in living cells. Therefore, a cell-free system derived from fission yeast nucleoplasmic extracts (YNPE) was employed[40]. We cultured fission yeast cells (Schizosaccharomyces pombe temperature-sensitive strain cdc25-22) at 37 $^{o}$C to arrest them in G2 phase and collected yeast cells of different phases at different time points after releasing at 25 $^{o}$C (see Fig. 2a, 2b and Fig. S5 in SI for details). To obtain YNPE, intact fission yeast nuclei were isolated from the cells synchronized in different cell cycle phases (Fig. S5 in SI) and crushed by high-speed centrifuge. YNPE contains all the soluble proteins in nuclei that support DNA replication, transcription, DNA damage repair and nucleosome assembly[40]. Most importantly, YNPE is mitochondrial-free and avoids cytoplasmic contamination including enzymes involved in fermentation, making it possible to manipulate the free energy level by adding exogenous ATP and ADP.

The H1 kinase assay is the standard method to measure Cdc2 activity by monitoring the amount of radioactive phosphate (from [γ-$^{32}$P]-ATPs) transferred to histone H1 proteins[29, 41]. It has been reported that Cdc2 activity remains low at the early G2 phase and increases gradually during G2 phase, finally rising to a peak from late G2 phase to M phase, but then dropping abruptly at the end of M phase[16, 21, 22, 29]. We measured Cdc2 activities in YNPEs derived from early G2, late G2 and M phase cells (Figs 2c and 2d). As shown in Fig. 2c, Cdc2 activity in late G2 phase is higher than that in early G2 phase. Cdc2 activity in M phase YNPE is 6 times higher than that of early G2 phase, which is consistent with previous reports in Xenopus egg extracts[30]. It has been well known that the major B-type cyclin, Cdc13, is necessary and sufficient to drive G2-M transition by promoting Cdc2 activity in S. pombe[29]. We then added various concentrations of exogenous purified Cdc13 protein to early G2 phase YNPE and measured Cdc2 activity afterwards. The low level of Cdc2 activity in early G2 phase YNPE rose with increasing Cdc13 concentrations, accordingly. When 0.9 μM Cdc13 was added, Cdc2 activity was elevated to the level of M phase YNPE (Fig. 2d). These results demonstrate that Cdc13 accumulation drives the activation of Cdc2 in the YNPE system, indicating, in turn, that the *in vitro* YNPE system faithfully simulates Cdc2 activation involving G2-M transition in the fission yeast cell cycle.

We measured the $[Pi]$ in YNPE by ion chromatography. The $[Pi]$ was 17.36±0.54 mM in early G2 phase YNPE and 17.40±0.48 mM in late G2 phase YNPE. The relative change of $[Pi]$ is smaller than those of $[ATP]$ (1 ~ 900 nM) and $[ADP]$ (80 ~ 300 μM) observed in our experiments (Table S2 and S3 in SI). Thus, $\Delta G$ changes induced by $[Pi]$ fluctuation can be ignored and we simply utilized $\ln([ATP]/[ADP])$ to represent $\Delta G$ in experiments (see details in Method). Meanwhile, in the YNPE system, no mitochondrion produces ATP *de novo*. After adding exogenous ATP, many biochemical reactions in YNPE converted ATP to ADP continuously, which caused $[ATP]$ and $[ADP]$ in YNPE to change constantly. Therefore, to study the kinetics of Cdc2 activity in YNPE, we measured real-time $[ATP]$, $[ADP]$ and Cdc2 activities at corresponding time points after adding exogenous ATP or ADP (see schematic in Fig. 2b).

**Bistability and hysteresis exist in ATP-driving Cdc2 activation in late G2 phase YNPE.**

To experimentally test the roles of ATP and $\Delta G$ in Cdc2 activation predicted in Fig.1, we first utilized late G2 phase YNPE, which contains a relatively high level of Cdc13, to measure Cdc2 activities after adding exogenous ATP and ADP. Exogenous ATP at 1 μM, 5 μM, 20 μM, 50 μM and 100 μM ATP was respectively added to late G2 phase YNPE; then Cdc2 activity, $[ATP]$ and $[ADP]$ were monitored at different time points. For all different concentrations of ATP, Cdc2 activity reached the highest level at 1.5 minutes after adding exogenous ATP, gradually decreasing thereafter (Fig. S6). Accordingly, in the following Cdc2 activation experiments promoted by adding exogenous ATP, we chose the measured Cdc2 activity at 1.5 minute to represent the level of Cdc2 activation in late G2 phase YNPE.

To understand how exogenous ATP increases Cdc2 activity in late G2 phase YNPE, a series of concentrations of ATP (0, 1, 5, 20, 50 and 100 μM) was supplemented to YNPEs. As is shown in Fig. 3a and Table S2 in SI, after adding 1 μM or 5 μM ATP, Cdc2 activities remained at a low level. However, when 20 μM ATP were added, Cdc2 activity increased to a level 3.67 times higher than that without adding exogenous ATP. Cdc2 activity at this level is in the range of M phase Cdc2 activity, according to Fig. 2c. However, Cdc2 activity did not increase correspondingly when 50 μM and 100 μM ATP were added to the late G2 phase YNPEs, respectively. On the contrary, the values were both less than that when 20 μM ATP were added (6.41 a.u.). This phenomenon could be attributed to the competition effect between radioactive ATPs and non-radioactive ATPs in the H1 kinase assay for Cdc2 measurement (see the experimental data and theoretical analysis in subsection 2.2 and Fig. S7 in SI). Taken together, these observations reveal that the addition of exogenous ATP boosts Cdc2 activity in late G2 phase YNPE. As mentioned above, this process for promoting Cdc2 activity by adding exogenous ATP is designated as the 'going-up' process.

If YNPE is incubated at 25 °C after adding exogenous ATP, we have experimentally found that ATP is consumed constantly in YNPE with the $[ATP]/[ADP]$ ratio decreasing over time (Table S2 and Fig. S8 in SI). Based on this observation, we sought to characterize the hypothetical M-G2 transition, namely the 'coming-down' process, under the ATP consuming circumstance. To accomplish this, 20 μM ATP were first added to late G2 phase YNPE to boost Cdc2 activity to M phase level. Aliquots of YNPE were then removed at different time points for Cdc2 activity assays and measurement of both $[ATP]$ and $[ADP]$. As shown in Fig. 3b and Table S2 in SI, during the first 5 minutes after adding ATP, the values of measured real-time $[ATP]$ are in the range of 152.6 ~ 24.0 nM in YNPE, and the corresponding Cdc2 activities remain at high state (6.41 ~ 4.93 a.u.). Between 30 and 50 minutes of incubation, however, the values of measured real-time $[ATP]$ decrease to 1 ~ 2 nM, and the corresponding Cdc2 activities decrease to 1.59 ~ 2.27 a.u. (low state). Interestingly, although ATP concentration and $\ln([ATP]/[ADP])$ drop abruptly in the first five minutes, Cdc2 remains at the high state. In Fig. 3c and Fig. S13a, we use $[ATP]$ and $\ln([ATP]/[ADP])$ as the X-Y axis and Cdc2 activity as the Z axis to illustrate the 'going-up' process (adding exogenous ATP, in red line) and the 'coming-down' process (consuming ATP in YNPE, in green line). More detailed analysis and discussion about the ATP-consuming process are also seen in the section entitled "ATP-consuming process and high phosphorylation energy level in the YNPE system" and section 1.4 in SI.

Since increasing the $[ADP]$ in YNPE is another way to decrease the $\Delta G$ level in the system, we added the mixture of 20 μM ATP and 0.1mM ADP to late G2 YNPE to produce a lower $\Delta G$ level than that observed in the 20 μM ATP experiments mentioned above. Interestingly, different from only adding 20 μM ATP to YNPE, in which the highest level of Cdc2 activity appeared at 1.5 minutes after adding exogenous ATP, Cdc2 activity reached the highest level at 3 minutes after adding the mixture of ATP and ADP (Fig. S9b). Therefore, to monitor the 'going-up' process in this circumstance, we chose the measured Cdc2 activity, $[ATP]$ and $[ADP]$ at the 3-minute time point to represent the system's status for the following experiments. As observed in our experiments that only added 20 μM ATP, 20 μM ATP +0.1 mM ADP obtained the highest Cdc2 activity, and Cdc2 activities decreased gradually with 50 μM ATP +0.1 mM ADP and 100 μM ATP +0.1 mM ADP in the 'going-up' process (Fig. 3d). The 'coming-down' process is shown in Fig. 3e, and the 'going-up' curve and 'coming-down' curve, as illustrated by $[ATP]$, $\ln([ATP]/[ADP])$, and Cdc2 activities, are shown in Fig. 3f and Fig. S13b. Similarly, in Figs. 3g ~ i and Fig. S13c, we carried out the experiments by adding 0.2 mM ADP and different concentrations of ATP in late G2 phase YNPE. If ADP concentration was raised to 0.2 mM, then the time point for highest Cdc2 activity was delayed by 6 minutes after adding 20μM ATP and 0.2 mM ADP (Fig. S9c). The results for the 'going-up' process and 'coming-down' process with exogenous 0.2 mM ADP and ATP are shown in Figs. 3g ~ i. Taken together, the observations imply that bistability and hysteresis exist in the activation of Cdc2 driven by ATP and $\Delta G$ in late G2 phase YNPE, which is consistent with our theoretical prediction in Fig. 1f.

From these experimental data, we notice the close $[ATP]$ concentrations (152.6 nM, 178.3 nM, and 182.4 nM) at which Cdc2 activity goes up to the highest level, even if different concentrations of exogenous ADP were added to late G2 phase YNPEs, namely $\Delta Gs$ in YNPE are different. Similarly, Cdc2 activity also comes down to the lowest level at close $[ATP]$ concentrations (between 1 nM and 2 nM) (Figs. 3c ~ 3i and Table S2 in SI). Compared with the U-shaped curve depicting the bistable regions of Cdc2 activation in our theoretical results (Fig. 1f), these experimental observations on the $[ATP]$ thresholds for Cdc2 activation are plotted with an imaginary U-shaped curve in Fig. 3j. When different concentrations of exogenous ADP are added to late G2 phase YNPEs, the thresholds of $[ATP]$ for Cdc2 activation in 'going-up' processes are close and independent of $\ln([ATP]/[ADP])$,while the thresholds of $[ATP]$ for Cdc2 inactivation in 'coming-down' processes (the hypothetical M-G2 transition) behave in a similar manner.

These results suggest that $\Delta G$ levels are relatively high in late G2 phase YNPE, corresponding to the ATP-constant U curve $\Delta G$ range in Fig. 1f. Even if more ADP were added to the late G2 phase YNPE system, it is notable that we could still not decrease $\Delta G$ level to a relevant lower range, so the lower $\Delta G$ range ($\Delta G < 46$ kJ/mol) in Fig. 1f was hard to achieve in the YNPE system.

**Bistability and hysteresis of Cdc2 activation were not observed in early G2 phase YNPE, but existed in early G2 phase YNPE supplemented with Cdc13**

To test the theoretical model for Cdc2 activation when Cdc13 concertation is low (Fig. 1e and 1f), we further investigated whether bistability and hysteresis exist in ATP-driving Cdc2 activation in early G2 phase YNPE, which contains a rather low level of Cdc13. For early G2 phase YNPE, Cdc2 activity reaches the highest level at the 1-minute time point after adding exogenous ATP (Fig. S10), so we chose the 1-minute time point to represent the system's status. After various concentrations of ATP (0, 1, 5, 20, 50, 100 and 200 μM) were added to early G2 phase YNPEs, we measured real-time $[ATP]$, $[ADP]$ and Cdc2 activity, as described above.

The 'going-up' process promoted by exogenous ATP is shown in Fig. 4a where Cdc2 activity rises with increasing exogenous ATP concentrations. The 'coming-down' process is shown in Fig. 4b where we measured real-time $[ATP]$, $[ADP]$ and Cdc2 activity at different time points in early G2 phase YNPE after 200 μM ATP were added. The $[ATP]$ and Cdc2 activity decreased gradually over time. In early G2 phase YNPE, Cdc2 activities of the 'going-up' curve and the 'coming-down' curve as a function of $[ATP]$ and $\ln([ATP]/[ADP])$ are illustrated in Fig. 4c, Fig. S13d and listed in Table S3 of SI.

Compared with the relevant results in late G2 phase YNPE in Fig. 3, it is likely that, in early G2 phase YNPE, no obvious bistability and hysteresis occur in the ATP-driving Cdc2 activation process. Cdc13 concentration in early G2 phase is well known to be lower than that in late G2 phase (Fig. S5). Together with the theoretical model (Fig. 1e and 1f), we infer that different Cdc13 concentrations should be a key point for bistability and hysteresis. Accordingly, we supplemented purified Cdc13 to the early G2 phase YNPE and examined the activation of Cdc2.

First, we supplemented early G2 phase YNPE with 0.3 μM Cdc13. The ATP-driving Cdc2 activation ('going-up') process is illustrated in Fig. 4d. At the 1-minute time point after 50 μM ATP addition to YNPE, the measured real-time $[ATP]$ is 244.3 nM, and the Cdc2 activity is about 2.57 a.u. Upon adding 100 μM ATP to YNPE, the 1-minute time point measured real-time $[ATP]$ is 398.7 nM, and the Cdc2 activity rises to the highest level (5.18 a.u.). The ATP-consuming 'coming-down' process is plotted in Fig. 4e where 100 μM ATP were added to YNPE and incubated at 25 °C. After 50 minutes of incubation, the measured $[ATP]$ was 4.5 nM, and the Cdc2 activity was 4.43 a.u. After 180 minutes of incubation, the measured $[ATP]$ dropped down to 2.7 nM, and Cdc2 activity deceased to the lowest level (~1.18 a.u.). The Cdc2 activities of the 'going-up' curve and the 'coming-down' curve as a function of $[ATP]$ and $\ln([ATP]/[ADP])$ are illustrated in Fig. 4f, Fig. S13e and listed in Table S3 of SI.

Next, we conducted similar experiments in early G2 phase YNPE by supplementing 0.45 μM purified Cdc13 (Figs. 4g ~ 4i and Table S3 in SI). In the 'going-up' process (Fig. 4g), 20 μM exogenous ATP drive Cdc2 activity to the highest level at the 1-minute time point after adding ATP to YNPE. In the 'coming-down' process (Fig. 4h), after 50 minutes of incubation, the measured real-time $[ATP]$ is 2.8 nM, and the Cdc2 activity is 5.96 a.u. After 180 minutes of incubation, the measured $[ATP]$ dropped down to 1.2 nM, and the Cdc2 activity remained at a high level (~4.49 a.u.). Cdc2 activities of the ‘going-up’ curve and ‘coming-down’ curve as a function of $[ATP]$ and $\ln([ATP]/[ADP])$ are illustrated in Fig. 4i and Fig. S13f.

In Fig. 4j, for early G2 phase YNPEs supplemented with 0.3 μM and 0.45 μM Cdc13, we plotted the $[ATP]$ thresholds for Cdc2 activation in the ‘going-up’ process and Cdc2 inactivation in the ‘coming-down’ process with an imaginary U-shaped curve to depict the bistability and hysteresis of Cdc2 activation. More results combined with late G2 phase YNPE supplemented with different concentrations of exogenous ADP are illustrated in Fig. S11. Furthermore, in Fig. 4k, when more exogenous Cdc13 is added to early G2 phase YNPE, the thresholds of $[ATP]$ in ‘going-up’ processes for Cdc2 activation decrease. In a similar way, the thresholds of $[ATP]$ for Cdc2 inactivation decrease in ‘coming-down’ processes (hypothetical M-G2 transition), as shown in Fig. S12.

Taken together, no bistability and hysteresis are observed in the process of ATP-driving Cdc2 activation in early G2 YNPE. However, if early G2 phase YNPE is supplemented with Cdc13 (0.3 μM or 0.45 μM), bistability and hysteresis can be observed in the ‘going-up’ and ‘coming-down’ Cdc2 activation processes. Meanwhile, based on our data, these results imply that Cdc13 and ATP synergistically drive Cdc2 activation.

**ATP-consuming process and high phosphorylation energy level in the YNPE system.**

In YNPE experiments, exogenous ATP is consumed over time, together with the changes of $[ATP]$, $[ADP]$ and the ratio of $[ATP]$ to $[ADP]$. Based on our mathematical model, we simulated and analyzed the ATP consumption process in Cdc2 activation. When high initial level of ATP (exogenous $[ATP]$= 7mM) is added to the system, our theoretical simulations show that $[ATP]$ and $\ln([ATP]/[ADP])$ decrease with time, but Cdc2 activity can be kept at a high level for a certain duration that corresponds to the activation of Cdc2 in M phase. The details are shown in Section 1.4 of SI (Fig.S3). In the biological experiments, we have verified the stability and hysteresis in the Cdc2 activation process in late G2 phase YNPE and early G2 phase YNPE with 0.3 μM and 0.45 μM Cdc13 YNPE, respectively (Fig.3 and 4).

We revisited our theoretical prediction in Fig. 1f. When $\Delta G$ level is in the ATP-constant U curve $\Delta G$ range, the thresholds of $[ATP]$ for Cdc2 activation in 'going-up' processes are almost independent of $\Delta G$, similar to thresholds of $[ATP]$ for Cdc2 inactivation in the 'coming-down' processes. Our results suggest that $\ln([ATP]/[ADP])$ and $\Delta G$ levels are relatively high in early G2 phase and late G2 phase YNPE.

In Fig. 4k, we combined the results about the thresholds of $[ATP]$ for Cdc2 activation and Cdc2 inactivation in both early G2 phase YNPE with exogenous

Cdc13 and late G2 phase YNPE. We noticed that the more Cdc13 the system has, the less ATP threshold level is needed for Cdc2 activation, and, similarly, the less ATP threshold level is required for the transition of the Cdc2 inactivation 'coming-back' process. These experimental observations are consistent with the theoretical prediction in Fig. 1g. In the late G2 phase YNPE system, we found that the ATP thresholds for 'going-up' and 'coming-down' processes are 160.0 nM and 1.5 nM, respectively. Interestingly, this suggests that the concentration of Cdc13 in the late G2 phase YNPE system is around 0.35 μM higher than that of Cdc13 in the early G2 phase YNPE system.

Overall, our data suggest that Cdc13 and ATP synergistically drive Cdc2 activation, and we conclude and confirm the presence of hysteresis in ATP-driving and $\Delta G$-driving Cdc2 activation, providing a stable and robust genetic switch for G2/M transition in the cell cycle process.

**Discussion**

From a physical standpoint, a living cell is an open system that continuously exchanges chemicals with its environment in the form of both high chemical potential 'food' and low chemical potential 'waste'. A significant portion of the exchanged chemicals (for example, glucose) are used to produce ATP for intracellular energy[3], while others are used for sensing environmental information (for example, endocrine hormones and pheromones)[4]. Several modeling studies have also revealed how nonequilibrium and ATP hydrolysis free energy govern cellular regulatory functions and information processes[11, 12, 13, 14, 15, 42, 43, 44, 45] whereby ATP hydrolysis free energy quantitatively depicts the nonequilibrium status of the living system.

In this study, from both theoretical analysis and *in vitro* experiment of fission yeast cell cycle process, we strongly support that ATP hydrolysis free energy and cellular ATP levels play an important role in G2-M transition. We showed that the CDK activation circuit in G2/M transition exhibits the expected ultrasensitive switch with bistability and hysteresis by saddle-node bifurcation only when the eukaryotic yeast cell is far from equilibrium with a suitable range of free energy (the free energy $\Delta G$ >34.6 kJ/mol) (Fig. 1c). Moreover, this saddle-node bifurcation is robust against relatively large changes and fluctuations of cyclin protein concentration, ATP level and phosphorylation energy in cells.

First, our theoretical analysis and results indicate that saddle-node bifurcation provides a bistable switch for G2-M transition in cell cycles where increasing ATP levels and Cdc13 concentrations in the cell turn 'on' G2/M transition under a certain level of ATP hydrolysis free energy $\Delta G$ (Fig. 1). Second, experiments in the YNPE system prove the existence of the bistable switch in fission yeast G2-M transition, leading to the inference that both ATP levels and Cdc13 concentrations serve as signals of Cdc2 activation *in vitro* (Fig. 2 and Fig. 3), and increasing cellular ATP level causes reduction of the Cdc13 threshold for G2-M transition (Fig. 4). In the YNPE system, ATP is continuously consumed with the decreasing ratio of ATP/ADP during experiments. Our experimental results in Fig. 3j and Fig. 4j suggest that the YNPE system is working at the higher ATP/ADP ratio or $\Delta G$ levels, which should be the upper range of the U-shaped curve in Fig. 1f. It is hard to explore the threshold of $\Delta G$ from bistability to monostability in the YNPE system.

The present work clearly suggests a nontrivial interplay among regulatory circuit function, stability and nonequilibrium conditions that deserves further investigation. In fission yeast G2-M transition, we showed that the U-shaped bistable region in the ATP-$\Delta G$ space (Fig. 1c and Fig. 1f) is determined by the CDK-Wee1-Cdc25 regulatory circuit and PdP cycles in Fig. 1a. In the budding yeast DNA replication checkpoint pathway, we have discussed the autophosphorylation circuit of kinase Rad53 with positive feedback where we found a knife-shaped bistable region in ATP-$\Delta G$ space[46]. Furthermore, we analyzed different regulatory circuits of PdP cycles, together with different shapes of the bistable region in ATP-$\Delta G$ space[47].

Our results also suggest the interaction between cellular metabolism and cell cycle regulation. We observed that G2-M transition requires a significant cellular ATP level and $\Delta G$ consumption. Accumulating evidence suggests that active Cdc2 enters mitochondria and promotes cellular ATP generation at the G2-M transition[48, 49]. Notably, in comparison with normal cells, mitochondrial respiration and glycolysis in cancer cells are altered to produce more ATP[50, 51]. Since tumor cells are in an actively dividing state, it is advantageous for them to adopt the most efficient strategy to ensure successful and ongoing cell cycles. In light of the metabolic costs and physical limitations of producing excess cyclin within cells, increasing cellular ATP level to reduce the required cyclin concentration threshold seems to be an optimal alternative approach for cancer cells. Meanwhile, recent studies have determined the key role of cellular ATP level in T cell activation, for example, CD8+ T effector cells and T helper 17 cells[52, 53]. Deficient ATP levels may result in inactivated or abnormal T cells that fail to control tumor development. Therefore, we rationalize the possibility of developing new immunotherapies based on controlling cellular ATP level in both cancer cells and immune cells.

More generally, living cells are far from equilibrium state with self-organization and dissipative structure[2]. For important cellular information and decision-making processes, the mechanism caused by saddle-node bifurcation should provide dynamic robustness against internal and environmental noises, which should be better than the transcritical bifurcation suggested by G. Nicolis and I. Prigogine[54]. This interplay of nonequilibrium and nonlinearity allows the execution of irreversible cellular processes, which should help us to understand the evolution of genetic circuits[16, 55, 56] and their application to develop synthetic genetic circuits and motifs.

## Method

### Strain and growth condition

Schizosaccharomyces pombe strain CDC25-22 (LD722 h+ leu1-32 his3-D1 cdc25-22) was used in this study. The cells were arrested at G2 phase when cultured at temperatures higher than 35 °C for 4 - 6 hours[57]. The arrested cells were released to continue normal cell cycle at 25 °C.

### Preparation of Yeast NucleoPlasmic Extracts (YNPE)

The preparation of YNPE was modified from reference[40]. Five-liter yeast cultures were grown at 25 °C to early-log phase in a fermentor (stirring speed 300 rpm, air flow rate 3 L/min). Cells were arrested at G2 phase at 37 °C for 5 h and released at 25 °C for an additional 120 minutes or 190 minutes (Fig. S5). After harvesting cells by centrifugation at 5000 rpm for 3 min at 4 °C, cells were washed twice with cold water and once with 40 mL YE1 buffer (1.2 M Sorbitol, 100 mM NaCl, 50 mM HEPES pH 7.6, 50 mM KCl, 2.5 mM MgCl2, 1 μM DTT, and 50 μg/mL Cycloheximide (Calbiochem)). Cells were resuspended with minimal volume (about 10 mL) of pre-chilled YE2 buffer (1.2 M Sorbitol, 100 mM NaCl, 50 mM HEPES pH 7.6, 50 mM KCl, 2.5 mM $MgCl_2$, 1 mM DTT, 20% glycerol, 5% polyvinylpyrrolidone 40 (molecular weight 40,000) (Sigma), 50 μg/mL Cycloheximide, 10 μg/mL Aprotinin/Leupeptin (Amresco), and 5 μg/mL Cytochalasin B (Millipore)) and frozen dropwise in liquid nitrogen. Frozen cells were ground using a SPEX 6870 freezer/Mill for 6 cycles of 2 min at a rate of 10 impacts per second. For each gram of ground cells, 4 mL pre-chilled YE2 buffer were added and mixed softly. Cell debris and residual whole cells were pelleted by 3800 g spinning for 5 minutes. The supernatants were then spun at 11,000 g for 10 minutes to pellet crude nuclei. The nuclei pellets were washed with YE3 (10 mM HEPES pH 7.6, 0.25 M Sucrose, 50 mM KCl, 2.5 mM $MgCl_2$, 1 mM DTT, 50 µg/mL Cycloheximide, 10 µg/mL Aprotinin/Leupeptin, and 5 µg/mL Cytochalasin B) twice. The nuclei were crushed at 260,000 g for 1 hour. After removing lipids on the top, the clear nucleoplasm was harvested. The yeast nucleoplasmic extracts (YNPE) were flash frozen in liquid $N_2$ for future use.

**Plasmid construction, protein expression and purification**

Full-length human histone H1 was expressed in *E. coli* BL21 (DE3) and purified as previously described[58]. The CDC13 gene was amplified by PCR from S. pombe gDNA with the forward primer:

5'-GAAGGATCCAATGACTACCCGTCGTTTAACTCG-3'

and the reverse primer:

5'-CGCGTCGACTTACCATTCTTCATCTTTCATGTCA-3'.

The CDC13 gene was cloned into the plasmid P-His-P1 to generate the plasmid P-His-P1-Cdc13. The Cdc13 protein was induced by 0.5mM isopropyl-β-d-thiogalactoside (IPTG) for 6 hours at 28 $^{o}$C and purified by HisTrap™ HP Sepharose column (GE Healthcare). The final protein concentration was determined by linear regression analysis on a protein gel using bovine serum albumin as a standard.

**Measuring Cdc2 activity, ATP and ADP concentrations**

Cdc2 activity was determined by histone H1 kinase assay as previously described [29, 40, 59, 60]. Histone H1 kinase assays were performed in a reaction system (50mM Hepes-KOH, pH 7.6, 25 mM ß-glycerol phosphate, 1 mM EDTA, 1 mM sodium orthovanadate, 1mM dithiothreitol, 10 mM $MgCl_2$, 1 μCi [$\gamma$-$^{32}$P]-ATP and 2 μg histone H1 in 20 μL) at 25 °C.

ATP concentration was measured by ATP assay kit (Beytone S0026B). The glycolysis method was used to determine the concentration of ADP in YNPE, as previously described[61, 62].

**Western blot**

The periodic oscillation of Cdc13 protein level in cell cycle was detected by anti-Cdc13 antibody (abcam, ab10873) (Fig. S5a). β-actin was used as loading control and detected by anti-β-actin (abmarking, ABM0132Po).

**Using $\ln([ATP]/[ADP])$ to indicate experimental $\Delta G$**

In the YNPE system, compared with manipulated $[ATP]$ (1 ~ 900 nM) and $[ADP]$ (80 ~ 300 μM), $[Pi]$ fluctuation will barely affect $\Delta G$. For example, as $[Pi]$ ranges from 16.82 mM to 17.90 mM in early phase YNPE, we supposed two experimental conditions: (1) $[ATP]_1 = 100nM$, $[ADP]_1 = 300\mu M$ and $[Pi]_1 = 16.82mM$ and (2) $[ATP]_2 = 900nM$, $[ADP]_2 = 100\mu M$ and $[Pi]_2 = 17.9mM$. The corresponding $\Delta G$ change ($\Delta G_1 - \Delta G_2$) is given by,

$$
\begin{aligned}
\Delta G_1 - \Delta G_2 &= RT\ln\left(\frac{[ATP]_1}{K_{eq}[ADP]_1[Pi]_1}\right) - RT\ln\left(\frac{[ATP]_2}{K_{eq}[ADP]_2[Pi]_2}\right) \\
&= RT\ln\left(\frac{[ATP]_1}{[ATP]_2}\right) - RT\ln\left(\frac{[ADP]_1}{[ADP]_2}\right) - RT\ln\left(\frac{[Pi]_1}{[Pi]_2}\right) \quad , \\
&\approx -2.20RT - 1.10RT + 0.0622RT \approx -3.24RT
\end{aligned}
$$

Where $\Delta G$ changes caused by $[ATP]$ and $[ADP]$ manipulations are ~20 times larger than $\Delta G$ change induced by $[Pi]$ fluctuation. Therefore, in our experiments, we ignored $[Pi]$ changes and used $\ln([ATP]/[ADP])$ to indicate $\Delta G$.

**Acknowledgement**

We thank Dr. Lilin Du for the gift of S. pombe strain CDC25-22, Dr. Xiangdong Li, Dr. Chao Tang and Dr. Kim Sneppen for helpful discussions. F.L. was supported by the National Natural Science Foundation of China (Grants No. 12174007 and 12090054), the National Key R&D Program in China (Grants No. 2018YFA0900200 and 2020YFA0906900), and the Li Foundation. Y.V.F. was supported by the National Natural Science Foundation of China (Grant No. 31970553 and 31371264), the National Key Research and Development Program of China (2021YFC2301000 and 2019YFA0905500), and the Newton Advanced Fellowship (NA140085) from the Royal Society.

## Figures

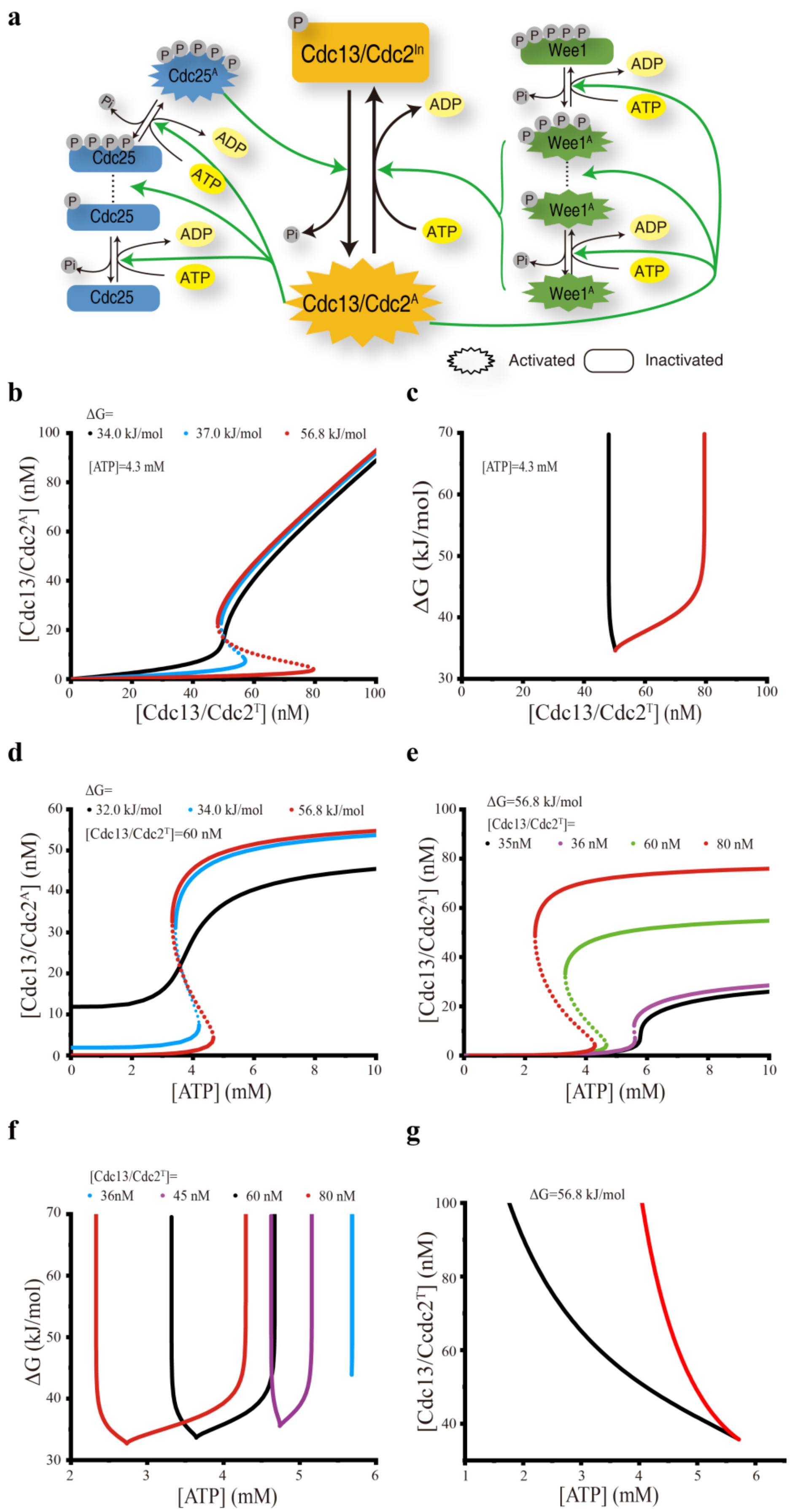

Figure 1| Regulatory circuit of fission yeast G2-M transition and theoretical model predicting bistability and hysteresis in Cdc2 activation controlled by cyclin, ATP and phosphorylation energy.

a, Reciprocal positive and double-negative feedback loops in the Cdc13/Cdc2-Wee1-Cdc25 circuit in the fission yeast cell cycle process. The PdP cycle of Cdc2/Cdc13 is at the center. Cdc2/Cdc13 complex is inactive when its Tyr15 is phosphorylated via active kinase Wee1, but this inactivation is reversed when dephosphorylation occurs owing to active phosphatase Cdc25. In turn, Cdc2/Cdc13 phosphorylates Cdc25 at multiple sites to activate Cdc25 and phosphorylates Wee1 at multiple sites to inactivate Wee1.

b, Modeling analysis predicts that the activity of Cdc13/Cdc2 exhibits bistability and hysteresis in response to total cyclin concentration ($[Cdc13/Cdc2^T]$) under different phosphorylation energy ($\Delta G$) levels.

c, A bistable region of Cdc13/Cdc2 activity in G2-M transition; U-shaped curve presents as having functions of both $[Cdc13/Cdc2^T]$ and $\Delta G$. The right branch in red corresponds to G2-M transition.

d and e, Modeling analysis predicts that the activity of Cdc13/Cdc2 exhibits bistability and hysteresis in response to ATP under different $\Delta G$ and $[Cdc13/Cdc2^T]$.

f, Bistable regions of Cdc13/Cdc2 activity in the G2-M system are determined by cellular ATP, $\Delta G$ and $[Cdc13/Cdc2^T]$. The right branch of each U-shaped curve corresponds to G2-M transition. When $[Cdc13/Cdc2^T]$=60 nM during the high $\Delta G$ range ($\Delta G > 45.7$ kJ/mol),ATP thresholds for Cdc2 activation basically remain constant.

g, Modeling analysis predicts that less ATP threshold for Cdc2 activation is needed for high level of $[Cdc13/Cdc2^T]$.

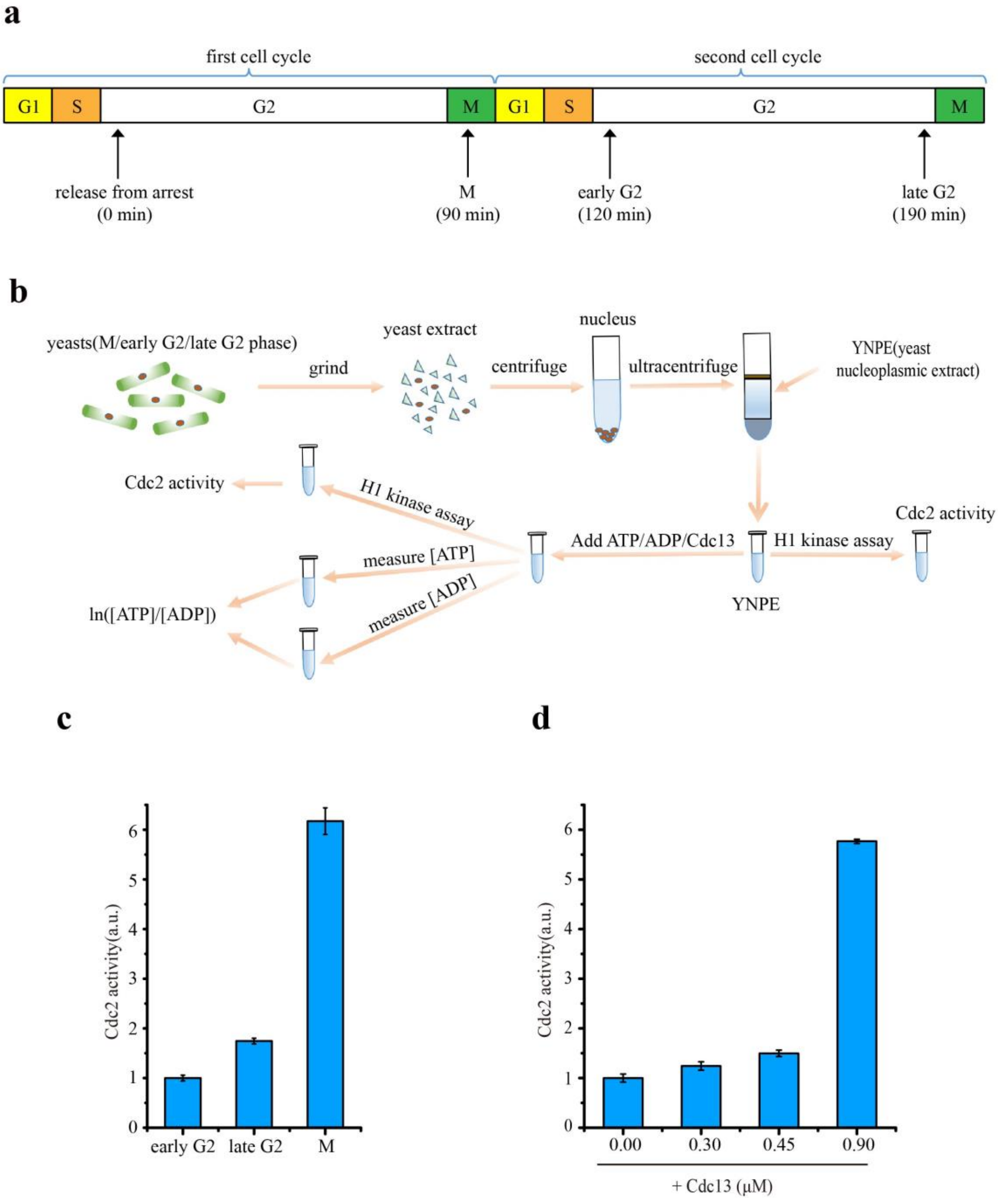

Figure 2| The yeast nucleoplasmic extracts (YNPE) system was used to study the activation of Cdc2 in fission yeast G2-M transition.

a, Schematic diagram of different cell cycle phases of fission yeast (M phase, early G2 and late G2 phase), which correspond to the different time points after release from arrested G2 phase.

b, Schematic diagram of preparation of YNPEs from cells in different cell cycle phase and the design of experiments. In real-time Cdc2 activity, ATP and ADP concentrations of the YNPE system were measured at the indicated different time points.

c, Cdc2 activities in early G2 phase, late G2 phase and M phase YNPEs. Values on the Y-axis indicate relative Cdc2 activity normalized to Cdc2 activity in early G2 phase YNPE. All error bars represent standard deviation of three independent experiments.

d, Cdc2 activity increased after adding various concentrations of exogenous Cdc13 in early G2 phase YNPEs. Same as (c), Cdc2 activities were normalized with Cdc2 activity in early G2 phase YNPE without exogenous Cdc13 addition. All error bars represent standard deviation of three independent experiments.

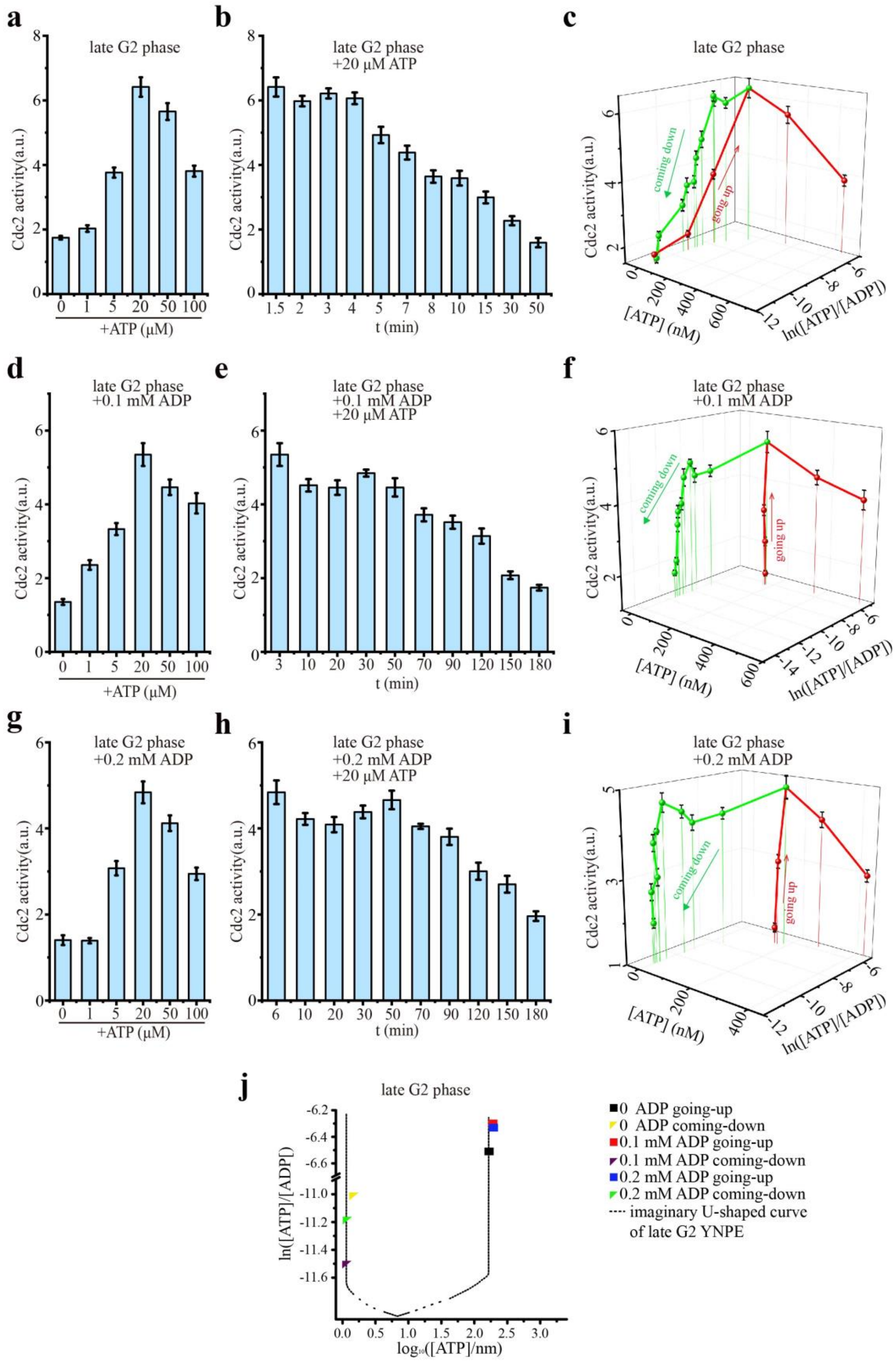

Figure 3| Bistability and hysteresis of Cdc2 activation in late G2 phase YNPE.

a~c, Bistability and hysteresis of Cd2 activity in ATP-driving and ATP-consuming processes in late G2 phase YNPE. (a) Exogenous ATP promotes Cdc2 activation, which is noted as the 'going-up' process. Cdc2 activities were measured at 1.5 minutes after exogenous ATP was added to late G2 phase YNPEs. (b) In the ATP-consuming process, late G2 phase YNPE was incubated at 25 ℃ after adding 20 μM exogenous ATP, and Cdc2 activity changed with time. This is noted as the 'coming-down' process. (c) Cdc2 activity of both 'going-up' (red) and 'coming-down' (green) processes plotted in the $[ATP]$ and $\ln([ATP]/[ADP])$ space.

d~f, Bistability and hysteresis of Cdc2 activity in ATP-driving and ATP-consuming processes in late G2 phase YNPE with 0.1 mM exogenous ADP. (d) The 'going-up' process. Exogenous ATP was supplemented to promote Cdc2 activation. Cdc2 activities were measured at 3 minutes after exogenous ATP and 0.1 mM ADP were added to late G2 phase YNPEs. (e) Cdc2 activity in the ATP-consuming and 'coming-down' process. Late G2 phase YNPE was incubated at 25 ℃ after adding 20 μM exogenous ATP and 0.1 mM ADP. (f) Cdc2 activity of (d) 'going-up' (red) and (f) 'coming-down' (green) processes plotted in the $[ATP]$ and $\ln([ATP]/[ADP])$ space.

g~i, Bistability and hysteresis of Cd2 activity can be observed in the ATP-driving and ATP-consuming processes in late G2 phase YNPE with 0.2 mM exogenous ADP. (g) The 'going-up' process. Exogenous ATP promotes Cdc2 activation.

Cdc2 activities were measured at 6 minutes after exogenous ATP and 0.2 mM ADP were added to late G2 phase YNPEs. (h) Cdc2 activity in the ATP-consuming and 'coming-down' process, and late G2 phase YNPE was incubated at 25 ℃ after adding 20 μM exogenous ATP and 0.2 mM ADP. (i) Cdc2 activity of (g) 'going-up' (red) and (h) 'coming-down' (green) processes plotted in the $[ATP]$ and $\ln([ATP]/[ADP])$ space.

j, In late G2 phase YNPE with different exogenous ADP, the $[ATP]$ thresholds that Cdc2 activity transits in 'going-up' and 'coming-down' curves are plotted in the $[ATP]$ and $\ln([ATP]/[ADP])$ space, together with the similar U-shaped curve in Fig. 1f depicting bistability and hysteresis of Cdc2 activation.

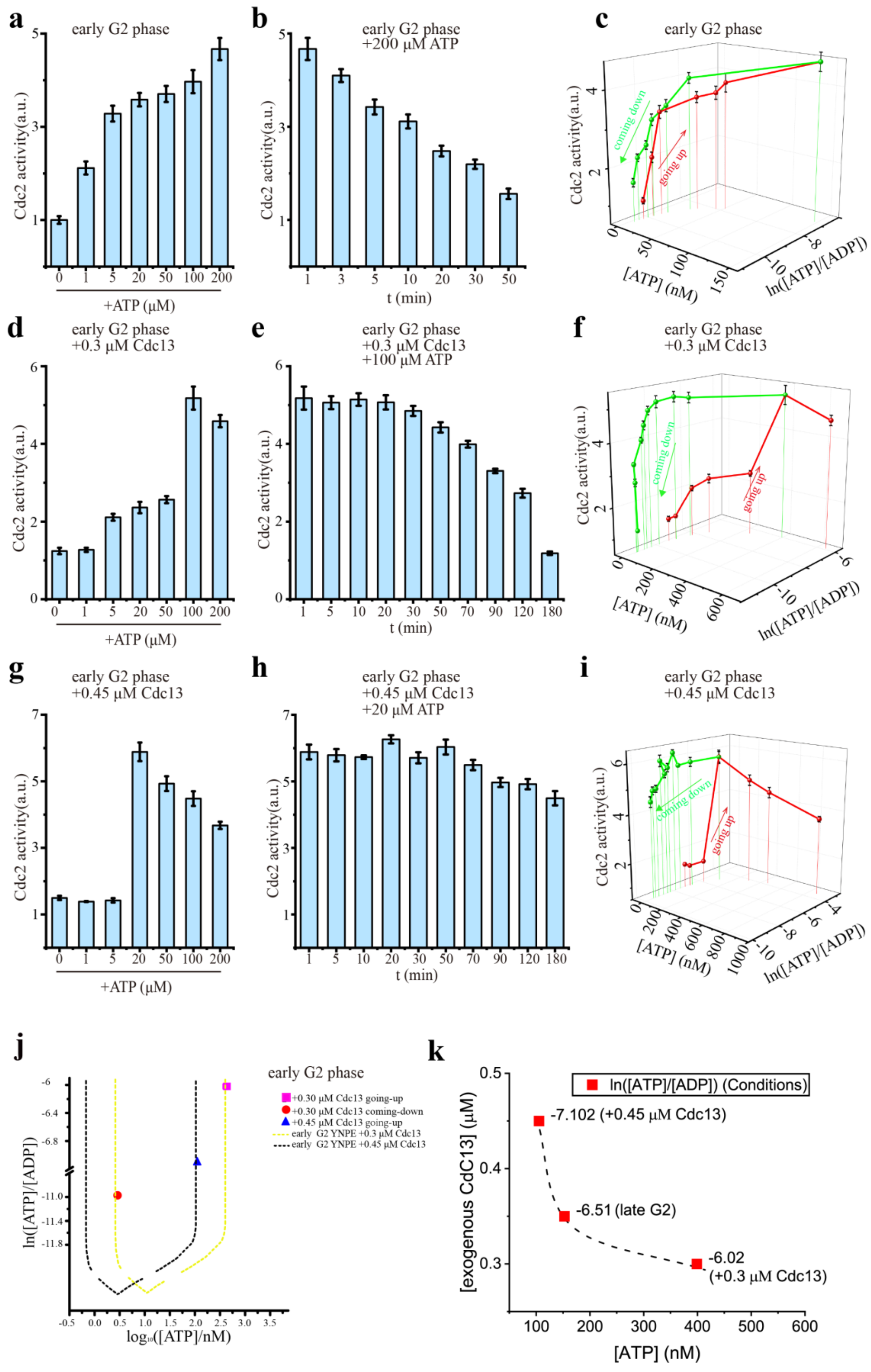

Figure 4| Cdc2 activation driven by ATP in early G2 phase YNPE supplemented

with different concentrations of exogenous Cdc13.

a~c. In early G2 phase YNPEs, no significant hysteresis of Cd2 activity can be observed in ATP-driving and ATP-consuming processes. (a) The ‘going-up’ process. Cdc2 activation was promoted by adding exogenous ATP where Cdc2 activities were measured at 1 minute after exogenous ATP was added to the early G2 phase YNPEs. (b) The ‘coming-down’ ATP-consuming process. YNPE was incubated at 25 ℃ after adding 200 μM exogenous ATP. Cdc2 activity decreases owing to ATP consumption with time. (c) Cdc2 activity of both ‘going-up’ (red) and the ‘coming-down’ (green) processes plotted in the $[ATP]$ and $\ln([ATP]/[ADP])$ space.

d~f. Hysteresis of Cdc2 activity can be observed in ATP-driving and ATP-consuming processes in early G2 phase YNPE with 0.3 μM exogenous Cdc13. (d) The ‘going-up’ process. Cdc2 activation was promoted by adding exogenous ATP where Cdc2 activities were measured at 1 minute after exogenous ATP was added to YNPEs. (e) ATP-consuming ‘coming-down’ process. YNPE was incubated at 25 ℃ after adding 100 μM exogenous ATP. Cdc2 activity decreases owing to ATP consumption with time. (f) Cdc2 activity of both ‘going-up’ (red) and ‘coming-down’ (green) processes plotted in the $[ATP]$ and $\ln([ATP]/[ADP])$ space.

g~i. Hysteresis of Cdc2 activity can be observed in the ATP-driving and ATP-consuming processes in early G2 phase YNPE with 0.45 μM exogenous Cdc13.

(g) The ‘going-up’ process. Cdc2 activation was promoted by adding exogenous ATP where Cdc2 activities were measured at 1 minute after exogenous ATP was added to YNPEs. (h) ATP-consuming ‘coming-down’ process. YNPE was incubated at 25 ℃ after adding 20 μM exogenous ATP. Cdc2 activity decreases owing to ATP consumption with time. (i) Cdc2 activity of both ‘going-up’ (red) and ‘coming-down’(green) processes plotted in the $[ATP]$ and $\ln([ATP]/[ADP])$ space.

j, In early G2 phase YNPE with different exogenous Cdc13, the $[ATP]$ thresholds that Cdc2 activity transits in ‘going-up’ and ‘coming-down’ curves are plotted in the $[ATP]$ and $\ln([ATP]/[ADP])$ space, together with the similar U-shaped curve in Fig. 1f depicting bistability and hysteresis of Cdc2 activation.

k, Combining the results of the ‘going-up’ process in late G2 phase YNPE and early G2 phase YNPE with 0.30 μM and 0.45 μM Cdc13 where we plot the ln([ATP]/[ADP]) in each condition. Concentration of Cdc13 in the system is inversely correlated with the value of ATP threshold required for Cdc2 activation. Concentration of Cdc13 in late G2 phase YNPE system is around 0.35 μM higher than the Cdc13 concentration in the early G2 phase YNPE system.